\documentclass[nohyper,12pt,letterpaper]{JHEP}
\usepackage{epsfig}
\newfont{\frak}{eufm10 scaled 1200}

\newfont{\Bbb}{msbm10 scaled 1200}     %instead of eusb10
\newcommand{\mathbb}[1]{\mbox{\Bbb #1}}
\DeclareSymbolFont{AMSa}{U}{msa}{m}{n}
\DeclareSymbolFont{AMSb}{U}{msb}{m}{n}
\let\Box\relax
\DeclareMathSymbol{\Box}{\mathord}{AMSa}{"03}

\def \eqn#1#2{\begin{equation}#2\label{#1}\end{equation}}

\title{ Holographic Cosmology}

\author{T. Banks \\
    Department of Physics and Astronomy - NHETC\\
    Piscataway, NJ 08540\\
    and\\
    Department of Physics, SCIPP\\
    University of California, Santa Cruz, CA 95064\\
E-mail: \email{banks@scipp.ucsc.edu}}

\author{ W. Fischler,  \\
      Department of Physics\\
      University of Texas, Austin, TX 78712\\
E-mail: \email{fischler@physics.utexas.edu}}

\abstract{We describe a cosmology of the very early universe,
based on the holographic principle of 't Hooft and Susskind. We
have described the initial state as a dense black hole fluid.
Here we present a mathematical model of this heuristic picture, as
well as a non-rigorous discussion of how a more normal universe
could evolve out of such a state. The gross features of the
cosmology depend on a few parameters, which cannot yet be
calculated from first principles.   For some range of these
parameters, microwave background fluctuations originate from
fluctuations in the black hole fluid, and have characteristics
different from those of most inflationary models.}

\keywords{Inflation, Cosmology}

\preprint{hepth{0405200}\\SCIPP-04-37\\UTTG-04-01}

\begin{document}

%%%%%%%%%%%%%%%%%%%%%%%%%%%%%%%%%%%%%%%%%%%%%%%%%%%%%%%%%%%%%%%%%%%%%%%%%%%%
%          Table of contents automatic !!!                                 %
%%%%%%%%%%%%%%%%%%%%%%%%%%%%%%%%%%%%%%%%%%%%%%%%%%%%%%%%%%%%%%%%%%%%%%%%%%%%

\section{The quantum mechanical origin of a $p=\rho$ universe}

Superstring Theory (ST) is our most successful attempt at
constructing a quantum theory of gravitation. It gives us a
variety of quantum theories of gravity in asymptotically flat and
anti-de Sitter space-times.  Gauge invariant observables are
correlation functions on the boundary.  All possible knowledge
about the interior is supposed to be holographically reconstructed
from this boundary data.

General principles indeed suggest that there are no gauge
invariant local observables in a theory of quantum gravity.
Precise verification of the  mathematical predictions of quantum
theory requires infinitely large measuring devices (whose quantum
fluctuations can be made arbitrarily small).   In a theory of
gravitation, those devices will have uncontrollable gravitational
interactions with the system they are supposed to measure, unless
they are moved to infinite spatial distances. The combination of
quantum mechanics and gravitation precludes the possibility of
infinitely precise measurements that are also localizable.

This conclusion is reinforced by the covariant entropy bound
\cite{fsbv}, which implies in particular that a finite causal
diamond in any Lorentzian space-time has finite entropy.  We will
interpret this as the entropy of the maximally uncertain density
matrix for observations done inside the causal diamond - the
Hilbert space of a causal diamond in quantum gravity, is finite
dimensional.   {\it A fortiori} such a system cannot make
arbitrarily precise measurements on itself, because of the
intrinsic limitations of quantum mechanics.   Thus, the
mathematical theory of such a region is, of necessity, ambiguous.
Different time evolution operators, which agree only up to the
maximal allowed precision of self-measurements of a finite system,
will give equally good descriptions of the physics.   We view the
argument of this paragraph as the quantum mechanical origin of the
diffeomorphism gauge ambiguity of classical general relativity.

We will attempt to build a general quantum theory of space-time on
these remarks.   First, we try to model an observer following a
time-like trajectory through space-time as a sequence of causal
diamonds.   Describe the trajectory by a sequence of overlapping
intervals ${\cal I}_n$, whose length increases with $n$.  For a
Big Bang cosmology it is convenient to choose intervals which all
begin on the initial singularity.  We will stick to this choice in
the present article.   The sequence of intervals defines a
sequence of causal diamonds, of increasing area\footnote{The area
of a causal diamond is the maximal area of $d - 2$ surfaces
obtained by foliating its boundary.}.  Our interpretation of the
covariant entropy bound then allows us to associate a Hilbert
space of fixed dimension with each interval.   Inversely, we can
make a {\it quantum mechanical definition of a localized observer}
as such a sequence of Hilbert spaces, of increasing dimension.

The nesting of causal diamonds is naturally associated with
causality.  This can be embedded into the quantum theory by
insisting that each Hilbert space in the sequence be a tensor
factor of the next one, ${\cal H}_{N+1} = {\cal H}_N \otimes {\cal
K}_N$.  Simplicity, and the considerations of
\cite{susyholoscreen} suggest that we choose a fixed ${\cal K}_N =
{\cal K}$, and we shall do so.  The dimension $d_{\cal K} \equiv
e^{l_{\cal K}}$ of ${\cal K}$ will not be terribly important in
this paper, but it is convenient to think of it as being the
representation space of a finite number of real fermionic
oscillators, $S_a$, so its dimension is a power of $2$. ${\cal K}$
represents those measurements in the $N + 1$st causal diamond,
which commute with all measurements in the $N$th diamond.

Each Hilbert space ${\cal H}_N$ is equipped with a sequence of
Hamiltonian operators, $H_N (k)$, for $ k = 1 \ldots N$.   These
represent time evolution over the discrete time intervals between
the future tips of the different causal diamonds.    Consistency
requires that

\eqn{consista}{H_N (k) = H_k (k) + V_N (k)}

where $V_N (k)$ depends only on those fermionic oscillators in
${\cal H}_N$ which are not in ${\cal H}_k$.

Consider two time-like trajectories in a Big Bang cosmology, and
choose a time slicing so that the causal diamonds defined by the
time $t$ along each trajectory have equal area, $A(t)$.  At each
time, we can inquire about the area $A_{D_1, D_2} (t)$ of the
maximal causal diamond that fits into the intersection between the
causal diamonds on our pair of trajectories.   We define two
trajectories to be nearest neighbors if, for all $t$ the
difference $A(t) - A_{D_1, D_2} (t) = 4 L_P^{d-2} l_{\cal K}$.
Starting with any trajectory we can find $2(d -1)$ independent
nearest neighbor trajectories.   Assuming the topology of spatial
slices is that of $R^{d-1}$, we can label a complete set of
nearest neighbor trajectories by points on a hypercubic lattice.

The quantum translation of this construction is a hypercubic
lattice of sequences of Hilbert spaces, ${\cal H}_N (\bf x)$. For
each pair of nearest neighbor points on the lattice we specify, at
each $N$, a tensor factor ${\cal O}_N (x, x+ \mu)$ of dimension
$e^{(N-1) l_{\cal K}}$of each Hilbert space.   These two factors
are identified. There is now a strong constraint on the sequence
of Hamiltonians in the two sequences of Hilbert spaces.   Each of
them induces a dynamical evolution on  ${\cal O}_N (x, x+ \mu)$
and these two sequences of evolution operators must agree.  This
constraint is both hard to understand, and hard to satisfy.  We
will present a particular solution of it in a moment.

Before doing so, we point out that geometrical intuition leads us
to expect a more elaborate set of similar constraints.   Indeed,
if we consider two points on the lattice of trajectories that are
separated by more than a single step, the causal diamonds of these
trajectories may have an intersection, and we can inquire what the
maximal area causal diamond, which fits inside it, is.  It is
easy to see that the answer to this question depends on the
classical space-time geometry.   Thus, in the quantum mechanics,
we identify a tensor factor ${\cal O}_N (\bf x, \bf y)$ of the
Hilbert spaces ${\cal H}_N (\bf x)$ and ${\cal H}_N (\bf y)$, but
when the two points are not nearest neighbors we do not specify
the dimension of this overlap space, except for the following set
of inequalities.   Consider any path on the lattice, between ${\bf
x}$ and ${\bf y}$ with a minimal number of steps. If ${\bf z}$ is
any intermediate point along this path, then ${\cal O}_N (\bf x,
\bf y)$ should be a tensor factor of both ${\cal O}_N (\bf x, \bf
z)$ and ${\cal O}_N (\bf z, \bf y)$. Again there will be
dynamical constraints relating time evolution in different
Hilbert spaces, requiring that they agree on all overlaps.

We conjecture that every solution of this complicated web of
consistency conditions will be a quantum cosmology.  That is, when
$N$ is large, we will be able to construct a Lorentzian geometry
from the data given by the quantum system, which satisfies
Einstein's equation with a stress tensor obeying the dominant
energy condition. At the moment, we have one example of how this
works.

This is the $p=\rho$ Friedmann-Robertson-Walker universe that we
have dubbed a dense black hole fluid.  Our heuristic
discussions\cite{holcosm1n3} of this system identified it as a
highly entropic fluid that somehow managed to be completely
homogeneous.  We also conjectured that its dynamics obeyed an
exact scaling symmetry.   The high entropy content suggested to us
a model with a time dependent Hamiltonian, which at each instant
is chosen from a certain random distribution.  This guess also
fits with old results \cite{bkl} which suggest that dynamics near
a Big Bang singularity is chaotic.   A time dependent, random
Hamiltonian will move the system throughout its Hilbert space,
maximizing the available entropy.   Our identification of
space-time properties with quantum properties is built in such a
way that the system will exactly saturate the covariant entropy
bound.

To understand the precise ensemble of random Hamiltonians which we
use, begin by writing the term quadratic in fermion operators:

\eqn{quad}{ H_N (N) = {1\over N} \sum_1^N S_a (n) h_{mn} (N) S_a
(m) \equiv {1\over N} H_{1+1} ,}

where $h (N)$ is chosen independently at each $N$ from the
standard Gaussian distribution of $N \times N$ anti-symmetric
matrices.  It is well known that for large $N$, the spectrum of
$H_{1+1}$ approaches that of a free fermion conformal field
theory, with a UV cutoff of order one, on an interval of length
$N$.   In fact, this behavior is very universal.  One can add
arbitrary higher order polynomials in the $S_a (n)$, which take
the form of integrals of translation covariant short ranged
interactions, without changing this behavior, as long as one
chooses some signs in the collection of quartic terms (so that the
marginal perturbations are marginally irrelevant).   For each $N$,
we make an independent choice of Hamiltonian.  Thus, although the
spectral density approaches a universal one, the basis in which
the Hamiltonian is diagonal changes randomly.

We now choose the overlaps in the following way: given two points
${\bf x}$ and ${\bf y}$, let $k({\bf x, y})$ be the minimum number
of lattice steps in a walk between them.   Choose ${\cal O}_N
({\bf x,y})$ to be ${\cal H}_{N - k({\bf x,y})} ({\bf x} {\rm or}
{\bf y}) $,  where the ``or" refers to the fact that we must
identify the overlap Hilbert space as a tensor factor of {\it
both} ${\cal H}_N ({\bf x})$ and  ${\cal H}_N ({\bf y})$.   If we
now prescribe the sequence of Hamiltonians to be the same for all
trajectories (all points on the lattice),   then all of the
consistency conditions for the dynamics are satisfied.  The system
is homogeneous on the lattice.

The overlap prescription defines a {\it causal distance function}
on the lattice.   Namely, for any $N$ and any ${\bf x}$ we can
define the base of the backward lightcone from the tip of the
causal diamond to be those points whose overlap with ${\cal H}_N
({\bf x})$ has dimension $d_{\cal K}$.   On the lattice this
defines a hypercube oriented at angle ${\pi \over 4}$ to all of
the axes.   However, if we insist that the geodesic distance
between nearest neaighbor trajectories be the same, then the
distance along each path to this boundary is the same. The
coordinate hypercube is mapped into a geometrical
sphere\footnote{For those reader's of the right age, it may be
useful to call this the {\it carpenter's ruler map}.}. Thus, our
geometry is isotropic around each point. Note furthermore that
the spatial geometry defined in this way will be flat. It is
homogeneous, isotropic, open and has no intrinsic length scale.

In fact, there are two other indirect indications that the spatial
geometry should be flat, if we make the hypothesis (to be verified
below) that we are describing an FRW geometry satisfying
Einstein's equations with a perfect fluid stress tensor.   Our
system was built to saturate the covariant entropy bound at all
times. In an open FRW universe, this is only possible if the
equation of state is $p = \rho$ and the spatial geometry is flat.
Furthermore, the physics of our system obviously obeys a scaling
symmetry.   An FRW universe with $p= w\rho$ and spatially flat
geometry, always has a conformal Killing vector\footnote{This is a
necessary, but not sufficient condition for the quantum physics of
such a universe to have a scaling symmetry.}.  However, a curved
spatial geometry introduces a natural scale (the spatial curvature
at some value of the energy density), and so cannot have such a
symmetry.   Below, we will give further evidence that the scaling
symmetry of the $1 + 1 $ dimensional CFT should be identified with
the conformal Killing symmetry of $p = \rho$ cosmology.

We now want to show that the scaling laws of the dense black hole
fluid cosmology are reproduced by our quantum system. The energy,
$<H_N (N)>$  of a typical state of our system at time $N$ scales
like $N^0$, while the entropy scales like $N$.   If, as indicated
above, the system is to be identified as a flat FRW cosmology,
then the relation between entropy and cosmological time scales
like $N \sim t^{d - 2}$.   $H_N (N)$ is the Hamiltonian for a
unit interval of entropy.   Thus

\eqn{DeltaN}{\Delta N H_N (N) \sim H_N (N) \sim \Delta t H(t),}

where $H(t)$ is the time dependent Hamiltonian which generates
translation in cosmological time.  Since $\Delta N \sim t^{d - 3}
\Delta t$, we conclude that the energy in a give causal diamond
scales like $N^{{d - 3}\over {d - 2}}$.   This is the scaling law
for the mass of a horizon filling black hole, as we suggested in
\cite{holcosm1n3}.

The area of the causal diamond scales like $N$, and since the
spatial geometry is flat, its volume scales like $N^{{d - 1}\over
{d - 2}}$.   Thus, the cosmological energy and entropy densities
scale like

\eqn{rhoscal}{\rho \sim N^{- {2\over {d - 2}}} \sim {1 \over {t^2}}}

\eqn{sigmascal}{\sigma \sim N^{- {1\over {d - 2}}} \sim {1\over
t}}

The scaling law $ {1\over {t^2}} $ for the energy density follows
from the Friedmann equation for a general single component
equation of state.    The relation $\sigma \sim \sqrt{\rho}$ is
characteristic of the $p = \rho$ fluid with non-vanishing entropy
density.  Here we have derived these behaviors from quantum
mechanics.

\section{\bf  Phenomenology of a $p=\rho$ universe}

The quantum mechanical construction described above, gives rise to
the geometry of a FRW universe  which saturates the entropy bound.
This mathematically well defined  universe, because of  the
second law of thermodynamics, cannot  develop into a universe
like ours with an interesting history as it already saturates the
entropy bound.

We have proposed a phenomenological modification of the previous
geometry, which results in a universe with all the features that
are presently experimentally observed. This phenomenology  also
predicts an exact scale invariant spectrum of fluctuations in the
microwave background with a sharp cutoff on a scale of the
present size of the horizon.

  The starting point of this phenomenology is a geometry that has less
than maximum entropy. The $p=\rho$ fluid, is assumed here to
occupy, not the totality, but a large fraction ${1\over \epsilon}$
of the volume at the initial time.

   In addition  space  is also permeated by an intricate network of
"normal regions" \footnote{A "normal region" is one that is filled
with a fluid with an equation of state $p=w\rho$, with $
-1/3<w<1$}. This network is chosen to maximize the entropy with
the condition that it  survives avoiding being reabsorbed by the
$p=\rho$ fluid. This requirement is not trivial, indeed a single
isolated bubble of radiation that smoothly fits in the $p=\rho$
fluid will inevitably shrink as can be seen by applying the
Israel junction conditions. Moreover, survivability of the
network of "normal regions" also implies that the fluctuations in
the energy density in the normal region be small  in order that
these regions not  form black holes  and end up merging with the
$p=\rho$ fluid.

  This setup solves the horizon and flatness problems. Indeed consider
first the universe that is homogeneously filled with just a
$p=\rho$ fluid. It was shown \cite{fs} that such a  HOMOGENEOUS
fluid, can saturate the entropy at all times if there is no
spatial curvature.
  Positive spatial curvature leads  to recollapse, whereas in
negatively curved space where the volume grows in time like the
area, the entropy bound can at best be saturated at one instant
of time after which the ratio of the entropy to the area
decreases.
  The presence of the network does not alter these conclusions because
as was argued earlier, the fluctuations in the energy density of
the network have to be small in order for it to survive.
  The argument about the spatial curvature in the normal region can be
refined.  In order to maximize the entropy in  the "normal
regions" and have them survive they should be initially filled
with black holes  whose sizes are smaller than their separations.
Negative curvature impedes the efficient filling  of the volume
with such black holes whereas positive curvature violates the
requirement of survivability.

Initially then, the normal regions are filled with a dilute gas of
black holes which are  adequately described by the equation of
state of non-relativistic matter. Subsequently these black holes
decay and the energy density in these "normal regions"  becomes
dominated by radiation. \footnote { It is worth noting that there
are two phases for a gas of black holes, one is the $p=\rho$
fluid and the other is the non-relativistic gas phase. The order
parameter is the ratio between the size of the black holes and
their separation.} On equal area time slices, the fractional
volume occupied by the "normal regions" grows faster than the
volume occupied by $p=\rho$ and eventually overtakes the latter.

\eqn{VoverV}{{V_{w={1\over 3}}\over V_{w=1}}\sim \epsilon
t^{1\over 2}}, where $\epsilon$ is the initial fraction of
volumes, and $V_w$ is the volume occupied by a fluid with
equation of state $p=w\rho$

Thus, at time $T={1\over {{\epsilon}^2}}$, the volume of the universe
becomes dominated by the "normal regions". In their expansion,
the "normal regions" encompass interstitial regions of $p=\rho$
fluid which then appear as large black holes in the "normal
regions". The distribution in size and location of these large
black holes is uniform, with small fluctuations inherited from
the fluctuations of the network of "normal regions" in the
primordial $p=\rho$ fluid. It is now worth reminding the reader
that a $p=\rho$ dominated universe possesses a scale invariance.
All spatially flat FRW cosmologies have a conformal Killing
vector. This however does not imply scale invariant physics in
general. For the special case of a $p=\rho$ universe, we
conjectured that the quantum dynamics of the fluid was scale
invariant. Above, we have described a model which has this
symmetry. \footnote{In more general FRW universes, the existence
of a locally measurable temperature tells the observer about the
existence of a scale.}.

We also showed in earlier work \cite{holcosm1n3} how this scale
invariance gets imprinted from the fluctuations in the energy
density of the network onto fluctuations in the energy density in
the normal part of the universe. The model predicts a
Harrison-Zeldovich spectrum for a finite range of scales.

A problem remains: the universe as described so far, has only
undergone subluminal expansion. Therefore the spectrum of scale
invariant perturbations cannot extend  to the present horizon. The
need then arises for a short burst of inflation in order to
stretch the scale invariant spectrum up to the size of the
present horizon. This can be easily achieved if the low energy
effective theory in the normal regions is one derrived from
superstring theory.  There will be moduli in the " low energy"
spectrum, which can potentially play the role of inflaton.

Let us now return to the era just after the large black holes
appeared in the "normal region".  It is clear, since we have just
made the transition from the dense black hole fluid, that black
holes will dominate the energy density for a time. Eventually the
black hole energy density redshifts , enabling the energy density
stored in the inflaton to take over, and the universe inflates.
This inflation however does not erase the fluctuations of the
energy density that was stored in the large black holes and their
decay products. The start of the inflationary era at a given
point, depends on the local black hole density.

The requirements that the scale invariant spectrum of fluctuations
extends from the present horizon size, down to a few decades from
that scale, that inflation starts at the end of the black hole
dominated era and that the reheating temperature be sufficient
for nucleo-synthesis implies as was shown in \cite{holcos 3.0},
that the number of  e-foldings , $N_e$

\eqn{Ne}{17\leq N_{e}\leq 41}

The reheat temperature in our phenomenology is relatively low
and  at most can reach $10^{8} GeV$.  In principle, the number of
e-foldings is predicted in terms of the microscopic parameters of
our model, but we do not yet know how to calculate them.   As a
consequence, there are two possible ways in which our model could
be compatible with the data.   In the first (which many inflation
theorists will find fine-tuned) the range of fluctuation scales
observable in the cosmic microwave background overlaps with the
range where fluctuations were generated during the $p=\rho$ era.
This predicts an exactly scale invariant spectrum with cut-offs
on both the upper and lower end.

The second is less exciting.   The late, low scale inflation must
generate the required fluctuations.  This probably is only
compatible with hybrid (rather than slow roll) inflation models,
and there are probably many constraints on the model following
from compatibility with the $p=\rho$ initial conditions. We have
not yet done a serious investigation of models that could fill
the bill.

\section{Conclusions}

Our heuristic discussion of the homogeneous dense black hole fluid
has now been replaced by a precise mathematical model.  One would
like to put the whole of our phenomenological cosmology on a
similarly rigorous footing.   From the practical point of view,
this is the only way we can imagine calculating the crucial
parameter $\epsilon$, which determines, among other things, the
range of scales over which we predict a scale invariant
fluctuation spectrum. Similarly, our current arguments tell us
that the amplitude of fluctuations is small, but do not allow us
to calculate it precisely. Nor do they allow us to estimate the
degree of Gaussianity of the fluctuation spectrum.

To proceed, we will have to find a quantum mechanical model for a
normal ({\it e.g.} radiation dominated )  FRW universe, and then
model the interface between a dense black hole fluid and a
radiation fluid.   This should help us to understand the
statistical properties of the most entropic universe which can
evolve into a normal region. Then we must understand how to merge
our cosmological model with string theory, to assess whether an
inflaton field with the right properties can arise.  We view this
as part of a larger project of understanding how conventional
approaches to string theory emerge from the causal diamond
approach to quantum gravity, which we have described.

Perhaps the least convincing part of our cosmology (in the first
scenario discussed above) is the ``conspiracy" between inflation
parameters and $p = \rho$ parameters which is necessary to ensure
that the scales for which we predict a scale invariant spectrum,
coincide with those which are observed in the cosmic microwave
background.   It is hard to assess whether this is fine tuning
without a better understanding of the fundamental physics. If a
complete model were found, in which this conspiracy followed from
the mathematics, it would be viewed as a great triumph, rather
than a deficiency of the model.  Nonetheless, it is worth asking
whether holographic cosmology can be grafted on to a more
conventional inflationary scenario,  since we feel rather
confident that the $p=\rho$ phase is a correct description of the
very earliest moments of a Big Bang cosmology.    The apparent
problem is that inflation must come after a relatively long
period of  dominance by a dense and then a dilute fluid of black
holes.    Thus the scale of inflation is forced to be low in any
holographic cosmology. For slow roll models without fine-tuning of
parameters,  a low scale of inflation cannot explain the
amplitude of density fluctuations.   There is a class of natural
hybrid models, based on supersymmetry\cite{gretal} which does not
have field theoretic fine tuning problems, and appears to be able
to fit the data with a low inflation scale.    It may be that, in
the end, holographic cosmology will simply serve as the prelude
to such an inflationary scenario.   The authors may perhaps be
excused for hoping for a more direct confrontation between
experiment and the fundamental structure of quantum gravity.

%=========================================================================
\acknowledgments We would like to thank G. Veneziano, R.
Brustein, M.Dine, R.Bousso, and L.Susskind for useful discussions.

The research of W. Fischler is based upon work supported by the
National Science Foundation under Grant No. 0071512. The research
of T. Banks was supported in part by DOE grant number
DE-FG03-92ER40689.

%=========================================================================

\end{document}